\documentclass[secnumarabic,amsmath,amssymb,floatfix, nofootinbib,tightenlines,nobibnotes, aps, twocolumn]{revtex4}
\usepackage{booktabs} 
\usepackage{array}    

\usepackage{graphicx}
\usepackage{slashed}
\usepackage[colorlinks=true,urlcolor=blue,linkcolor=blue, citecolor=blue]{hyperref}

\newcolumntype{C}{>{$}c<{$}} 
\newcolumntype{L}{>{$}l<{$}} 

\newcommand{\abs}[1]{\left| #1 \right|}

\renewcommand{\epsilon}{\varepsilon}
\renewcommand{\phi}{\varphi}
\renewcommand{\Psi}{\varPsi}

\begin{document}
	
\title{Beyond Poincaré Stresses: A Modern Quantum Field Theory Take on Hydrogen’s Electromagnetic Mass}
\author{Qasem Exirifard}
\email{qexirifa@uottawa.ca}

\author{Alessio D'Errico}
\email{aderrico@uottawa.ca}

\author{Ebrahim Karimi}
\email{ekarimi@uottawa.ca}

\affiliation{Nexus for Quantum Technologies, University of Ottawa, 25 Templeton Street, Ottawa, Ontario, K1N 6N5 Canada}

\begin{abstract}
We revisit the longstanding electromagnetic mass problem from a modern quantum field theory perspective. Focusing on a system of two widely separated hydrogen atoms---one in an excited \(nS\) state and the other in the ground \(1S\) state---we isolate the electromagnetic contribution to the electron’s total linear momentum by comparing the full energy-momentum tensor with the predictions of a point-like bound state model. Our analysis reveals that the leading perturbative correction introduces a factor \(4/3\), which, along with subsequent corrections, indicates that the effective electromagnetic mass deviates from the conventional relation \(E/c^2\). This discrepancy is attributed to the intrinsic nonlocality of the electromagnetic field, rather than to additional compensating mechanisms such as Poincaré stresses. We further contrast our quantum field theory results with the highly accurate predictions of the Schrödinger equation, which, despite neglecting higher-order terms, achieves an average error on the order of \(10^{-5}\%\). Attempts to improve this accuracy via perturbative inclusion of the self-interaction of the electron's wave function instead increase the error, prompting a re-examination of the underlying perturbative assumptions. Our findings suggest that a non-perturbative treatment of the tree-level action may be required to fully capture the dynamics of bound states in quantum field theory.
\end{abstract}
\date{\today}
    
\maketitle

J.J.\ Thomson’s early work~\cite{Thomson1897} revealed that the linear momentum of an electron’s electromagnetic field is given by
\begin{equation}
\vec{p}_{\text{field}} = \frac{4}{3}\frac{E}{c^2}\,\vec{v},
\end{equation}
where \(E\) is the electron’s rest electromagnetic energy, \(\vec{v}\) its velocity, and \(c\) denotes the speed of light in vacuum. This result implies an effective electromagnetic mass of
\[
m_{\text{em}} = \frac{4}{3}\frac{E}{c^2},
\]
which appears to conflict with Einstein's celebrated relation \(m = {E}/{c^2}\)~\cite{Einstein:11,Note}. This ``4/3 problem'' troubled early electron models in which the electron was treated as a rigid charged sphere~\cite{Abraham1903,Lorentz1909}. The discrepancy was emphasized by Born~\cite{Born1920} and Laue~\cite{Laue1911}. Fermi~\cite{Fermi1923} later argued that the rigid-body assumption was at fault, while Poincaré~\cite{Poincare1905} introduced nonelectromagnetic ``Poincaré stresses'' to stabilize the charged sphere.

Numerous attempts have been made to resolve this issue. For example, Feynman~\cite{Feynman1964} argued that proper treatment of Poincaré stresses eliminates the discrepancy, and Rohrlich~\cite{Rohrlich1965} along with Schwinger~\cite{Schwinger1973} proposed that the Abraham--Lorentz definitions of energy and momentum are not relativistically invariant and must be reformulated. Becker~\cite{Becker1964} and Morozov~\cite{Morozov1971} attributed the 4/3 factor to additional elastic potential energy.

In the modern Standard Model, the electron is regarded as a point-like fundamental particle, making additional contributions from Poincaré stresses or elastic energy unnecessary. However, the concept of electromagnetic mass remains of theoretical interest. In this work, we revisit electromagnetic mass from a contemporary quantum field theoretic perspective, with the aim of addressing this long-standing issue in a new context. Whereas our previous analysis~\cite{2024arXiv240500071E} focused on a free electron and established a framework in which localization leads to nonpoint-like behavior, here we extend the investigation to electrons bound in hydrogen atoms.

To this end, we consider a system composed of two widely separated hydrogen atoms: one in an excited \(nS\) state moving to the right with velocity \(v\), and the other in its ground \(1S\) state moving to the left with the same speed. Assuming the excited state is metastable, or that the time scale is short compared to its radiative lifetime, we neglect spontaneous emission and treat the electromagnetic field as approximately static (electrostatic binding). The large separation ensures negligible overlap between the electromagnetic fields of the atoms. Furthermore, the spherically symmetric electron distribution does not affect the proton’s internal quark dynamics, and the protons’ momenta cancel by symmetry. This configuration isolates the contribution of the electron to the total linear momentum, enabling a re-examination of its electromagnetic mass in the bound-state regime.

We extend our investigation by employing a quantum field theoretic approach to compute the system’s linear momentum via the energy–momentum tensor,
\begin{equation}\label{eq:momentum_tensor}
{p}_i = \int d^3x\,\Bigl(T^1_{0i}+T^2_{0i}\Bigr),
\end{equation}
where \(T^1_{\mu\nu}\) and \(T^2_{\mu\nu}\) are the energy–momentum tensors of the first and second atoms, respectively. Our goal is to compare this momentum with that predicted by the relativistic formula for point-like bound states,
\begin{equation}\label{eq:relativistic_momentum}
{p}_i = -\frac{{\cal E}_n - {\cal E}_1}{c^2}\,\gamma\,{v}_i,
\end{equation}
where 
\({\cal E}_n\) is the bound-state energy of the excited \(nS\) state, \({\cal E}_1\) is the ground-state energy of the \(1S\) state, and \(\gamma = \frac{1}{\sqrt{1-v^2/c^2}}\) is the Lorentz factor. The negative sign in \eqref{eq:relativistic_momentum} reflects our convention that the excited atom moves in the $+v$ direction while the ground-state atom moves in the $-v$ direction, so the net momentum points toward the excited atom’s motion. Any discrepancy between these two approaches must be attributed to deviation from a point-like particle in the electromagnetic field.

It is worth noting that our analysis is based on several simplifying assumptions. We model the electron as a spinless complex scalar field and restrict the electromagnetic interaction to its electrostatic component. While this neglects relativistic effects and spin contributions, previous fully relativistic analyses in classical electrodynamics \cite{Rohrlich1966} encountered the same 4/3 problem. This suggests that including spin and full relativistic corrections (for instance, through the Dirac equation) would not remove the discrepancy arising from the extended nature of the electromagnetic field. In standard quantum mechanics, the electron mass is treated as an empirical input that has already absorbed self-energy effects via renormalization. Hence, the remarkable accuracy of the Schrödinger equation is due to the fact that it effectively works with a renormalized mass, and additional explicit self-interaction corrections would lead to double counting.

The remainder of this paper is organized as follows: In Section~\ref{sec:QFT} we review the path integral approach to the quantum field theory of the hydrogen atom, modeling it as a bound state of an electron, a proton, and photons. We postulate that the renormalizability of the theory ensures the correct identification of all spectral features once the energy differences between any two states are determined. Subsequently, we truncate the action to zero-loop order.
    
In Section~\ref{sec:HydrogenModel} we describe our hydrogen atom model. In Section~\ref{sec:ElectronModel} we introduce a simplified electron model that neglects spin by representing the electron as a complex scalar field \(\phi\) with mass \(m\) and charge \(q\). The electron interacts with the electromagnetic field \(A_\mu\) via the gauge-covariant derivative. Assuming a flat spacetime and focusing on static, spherically symmetric solutions, we take the gauge field to be purely electrostatic, \(A_\mu=(V,0,0,0)\). In a low-energy approximation, the effective Lagrangian reduces to a form that yields a time-independent Schrödinger equation for the electron’s wave function \(\psi\). The electron–proton interaction is captured by a set of nonlinear, nonlocal partial differential equations, which we solve iteratively using perturbation theory. The zeroth-order solution describes a hydrogen-like atom, while subsequent corrections refine the potential and energy levels. A detailed analysis of the lowest energy states is presented.
        
In Section~\ref{sec:ProtonModel} we address the proton. Although protons are composed of three quarks bound by strong and weak forces, their internal structure has a negligible effect for our purposes. Thus, we model the proton as a uniform sphere of radius \(a\) with a homogeneous isotropic energy distribution (characteristic of a perfect fluid). The electric charge is evenly distributed on the surface of the sphere, ensuring a balance between the internal pressure and the outward electric forces.
    
In Section~\ref{sec:EMTensor} we compute the energy–momentum tensor of the electromagnetic field, the electron, and the proton in the rest frame of the hydrogen atom.
        
In Section~\ref{sec:Spectra} we determine the total mass of the hydrogen atom, defined as the spatial integral of the \(T_{00}\) component of the total energy–momentum tensor. We compute the mass difference between the excited \(nS\) state and the ground \(1S\) state, normalized by the mass difference between the \(2S\) and \(1S\) states. These theoretical predictions are compared with the experimental data, yielding an average error of \((0.13\pm0.04)\%\).

In Section~\ref{sec:LinearMomentum}, we compute the linear momentum of a moving hydrogen atom by performing a Lorentz boost on the energy-momentum tensor \(T_{\mu\nu}\) and integrating \(T_{0i}\) over space in the boosted frame. Our analysis reveals a discrepancy between the linear momentum derived from quantum field theory and that obtained by modeling the hydrogen atom as a point-like particle. Specifically, the leading-order perturbative calculation yields a factor of \(4/3\). We also evaluate the subleading corrections to this factor. This mismatch demonstrates that the discrepancy in the linear momentum of the electromagnetic field arises from the inherent nonlocality of the field configuration, rather than from any need to introduce Poincaré stresses or other compensating forms of energy.

In Section~\ref{sec:Conclusion} we present our final remarks and a discussion.

\section{Renormalized Quantum Field Theory}
\label{sec:QFT}
We model the hydrogen atom as a bound state of three quantum fields: the electron represented by \( e \), the proton represented by \( p \), and the photon represented by \( \gamma_\mu \). These fields are governed by a three-field action \( S = S[\gamma_\mu, e, p] \). In the path integral formulation of quantum field theory, the Feynman path integral is defined as:
\begin{eqnarray}
    Z&=&\int \mathcal{D}\gamma_\mu \, \mathcal{D}e \, \mathcal{D}p\,  
    e^{i S+i\int d^4x \, \left( J^\mu \gamma_\mu + J_e e + J_p p \right)},
\end{eqnarray}
where \( J^\mu \), \( J_e \), and \( J_p \) are external sources coupled to the photon, electron, and proton fields, respectively, and $Z$ is a functional of \(J^\mu, J_e, J_p\). This path integral \( Z \) encodes all the correlation functions of the theory, allowing the calculation of physical observables through functional differentiation with respect to the sources. The effective action represented by \( \Gamma[A_\mu, \phi_e, \phi_p] \) is defined by performing a Legendre transformation of the generating functional \( Z \). Specifically, it is given by
\begin{eqnarray}
    \Gamma &=& \ln Z-\int d^4x \, \left( J^\mu A_\mu + J_e \phi_e + J_p \phi_p \right),
\end{eqnarray}
where the classical fields \( A_\mu \), \( \phi_e \), and \( \phi_p \) are defined as the expectation values of the quantum fields in the presence of the sources:
\begin{eqnarray}
    A_\mu(x) &=& \frac{\boldsymbol{\delta}}{\delta J^\mu(x)} \ln Z, \\
    \phi_e(x) &=& \frac{\boldsymbol{\delta}}{\delta J_e(x)} \ln Z, \\
    \phi_p(x) &=& \frac{\boldsymbol{\delta}}{\delta J_p(x)} \ln Z.
\end{eqnarray}
Notice that $\Gamma$ is considered functional of $[A_\mu, \phi_e, \phi_p]$.
Here, \( A_\mu \) represents the classical electromagnetic potential, while \( \phi_e \) and \( \phi_p \) correspond to the classical electron and proton fields, respectively. The effective action \( \Gamma \) thus encapsulates the quantum corrections to the classical action \( S \) and serves as the generating functional for one-particle irreducible (1PI) correlation functions. 

The effective action of a perturbatively renormalizable theory can be expressed as a loop expansion series in powers of \(\hbar\). This expansion systematically incorporates quantum corrections at different loop levels. The loop expansion of the effective action \( \Gamma \) is given by
\begin{equation}
    \Gamma = S_0 + \hbar S_1 + \hbar^2 S_2 + \cdots,
\end{equation}
where:
\begin{itemize}
    \item \( S_0 \) is the tree-level action, obtained by substituting the quantum fields \( \gamma_\mu \), \( e \), and \( p \) with the classical fields \( A_\mu \), \( \phi_e \), and \( \phi_p \) in the original action \( S \). This represents the classical dynamics without any quantum corrections.
    \item \( S_1 \) is the one-loop correction, accounting for quantum fluctuations at the first order beyond the classical approximation.
    \item \( S_2 \) is the two-loop correction, incorporating higher-order quantum effects.
    \item The series continues with higher-order loop corrections \( S_3, S_4, \ldots \), each multiplied by increasing powers of \(\hbar\).
\end{itemize}
The effective action $\Gamma$ also plays a crucial role in defining physical quantities. For a free electron, these quantities include the charge of the electron and its mass. In the context of quantum field theory applied to the hydrogen atom, the physical quantities are represented by the energy levels and spectral lines observed in the atom's spectra. Specifically, the physical quantities of the hydrogen atom are the discrete energy differences that manifest themselves as spectral lines. \textbf{We postulate that the renormalizability of the theory ensures the correct identification of all spectral features once the energy difference between any two states is determined}. This postulate leads us to compare the measured value of \eqref{NistReport} with the theoretical calculation given in \eqref{Entheory}.

In the following, we develop a quantum field theory description of the hydrogen atom by truncating the theory at tree level (i.e., zero-loop order). In the hydrogen atom’s rest frame, we define its energy as the spatial integral over all space of the  $00$-component of the total energy-momentum tensor. Similarly, when the hydrogen atom is in motion, its linear momentum in direction $i$, is defined as the spatial integral over all space of the $0i$-components of the energy-momentum tensor. This framework allows us to theoretically compute the energy states and linear momentum of the hydrogen atom at tree level, thereby extending the analysis beyond the conventional Schrödinger equation approximation.

\section{Modeling the hydrogen atom}
\label{sec:HydrogenModel}
To study the spectra of the hydrogen atom, we model it as a system composed of an electron, a proton, and photons. Each component is detailed in the following subsections. 

\subsection{Modeling the electron and photons}
\label{sec:ElectronModel}
In our analysis, the intrinsic angular momentum of the electron, commonly referred to as spin, is not considered relevant. Consequently, we choose to model the electron using a complex scalar field, denoted as $\phi$. This scalar representation simplifies our mathematical framework while capturing essential properties. Specifically, the field $\phi$ is assigned a mass $m$ and is characterized by an electric charge of $q$  which interacts with the electromagnetic ($U(1)$ gauge) field $A_\mu$.

The  tree-level effective action of the scalar electron and the electromagnetic field, therefore, is given by:
\begin{eqnarray}
    \Gamma[\phi, A_\mu] &=& \int d^4 x \sqrt{-\det g} ({\cal L}_e+{\cal L}_\gamma)\\
    {\cal L}_e &=& g^{\mu\nu} (D_\mu \phi)^* D_\nu \phi - m^2 \phi \phi^*\\
    \label{L_gamma}
    {\cal L}_\gamma &=& - \frac{1}{4} F_{\mu\nu} F_{\mu' \nu'} g^{\mu\mu'} g^{\nu\nu'} - J^\mu A_\mu
\end{eqnarray}
where $D_\mu \phi = \partial_\mu \phi + i q A_\mu \phi$ denotes the gauge-covariant derivative, $F_{\mu\nu}= \partial_\mu A_\nu - \partial_\nu A_\mu$ is the field strength of $A_\mu$, $J^\mu$ is the background electromagnetic current produced by the proton, $g_{\mu\nu}$ is the metric, $g^{\mu\nu}$ is the inverse of the metric, and $\det g$ is the determinant of the metric. We consider flat spacetime geometry where the line element can be represented by:
\begin{equation}
    ds^2 = g_{\mu\nu} dx^\mu dx^\nu =  dt^2 - dr^2 - r^2 d\theta^2 - r^2 \sin^2\theta d\phi^2.
\end{equation}
Notice we work in the units of $c=\hbar=1$. In this study we are interested in the static spherical solutions where the gauge field is given by
\begin{equation}
    A_\mu = (V, 0, 0, 0) \label{A_mu},
\end{equation}
where $V=V(r)$, and the background current is produced by the proton charge distribution which for the purpose of electron's dynamics can be approximated by 
\begin{eqnarray}
J^\mu=(\rho_p, 0,0,0),    
\end{eqnarray}
where $\rho_p$ is the charge density of the proton. We assume that the electric charge is uniformly distributed on a sphere with radius $a$: 
\begin{eqnarray}
    \rho_p= -\frac{q}{4\pi r^2} \delta(r-a),
\end{eqnarray}
where $a$ is the radius of the proton, and the charge of proton is ``$-q$". Here we study spherical solutions where the wave function of the electron is given by 
\begin{eqnarray}
\label{phi_sol}
    \phi = \frac{1}{\sqrt{2m}} \psi(r) e^{ - i (E+m) t},
\end{eqnarray}
where $E$ and $m$ respectively encode the binding and rest  energies of the electron. The effective Lagrangian of $\psi$ than can be approximated to 
\begin{eqnarray}
    {\cal L}_e =  E \abs{\psi}^2 - q V \abs{\psi}^2 -\frac{\abs{\partial_r \psi}^2}{2m},
\end{eqnarray}
where the following approximation are implemented:
\begin{itemize}
    \item only linear terms in $E$ and $V$ are kept,
    \item $\abs{q V}\ll m$,
    \item $|E|\ll m$.
\end{itemize}
We refer to the above conditions as the low energy approximation. Utilizing ${\cal L}_e$ into the action, and noticing $\sqrt{-\det g} =r^2$, and taking the functional variation with respect to $\psi$ results to the equation of motion for $\psi$:
\begin{equation}
    \label{sch1}
    \left(- \frac{\partial_r (r^2\partial_r)}{2m r^2} + q V\right) \psi = E \psi,
\end{equation}
which is the time independent Schr\"odinger equation for spherical solutions. So this formulation effectively describes the quantum mechanical behavior of the electron under the approximation of non-relativistic energies. 

The functional variation of the Lagrangian \(\mathcal{L}\) with respect to the gauge field \(A_\mu\) leads to the equations of motion for \(A_\mu\). By focusing on the static spherical solutions in the low-energy regime, we derive the following equation:
\begin{equation}
\label{v_eq1}
\nabla^2 V = -(\rho_p + q |\psi|^2).
\end{equation}
Equations \eqref{sch1} and \eqref{v_eq1} constitute a system of nonlinear, non-local partial differential equations. In the following of this section we consider that electron sees the proton as a point like particle and use the approximation of $a=0$.  In our analysis, we treat the electron's contribution to the potential as a perturbative element and employ a perturbative approach as follows:
\begin{eqnarray}
    \left(- \frac{\nabla^2}{2m} + q V\right) \psi &=& i \partial_t \psi, \label{perturb_sch} \\
    \nabla^2 V &=& q \delta^3(r) - \epsilon q |\psi|^2 + O(\epsilon^2) \label{perturb_pot}
\end{eqnarray}
where the parameter \(\epsilon\) quantifies the perturbation strength. When \(\epsilon=0\), the system reduces to the idealized scenario of a point-like electron interacting solely with an external Coulomb field (i.e., the standard Schrödinger equation). Setting \(\epsilon=1\) recovers the full electromagnetic self-interaction. This parametrization permits an iterative solution: starting with the dominant contributions and subsequently incorporating higher-order corrections in \(\epsilon\). In summary, treating \(\epsilon\) as a perturbative parameter is postulated to enhance the predictive accuracy of our model by systematically incorporating self-field effects.

The perturbative analysis necessitates the expansion of the wave function, energy, and potential into perturbative series as follows:
\begin{figure}[t]
    \centering
    \includegraphics[width=.90\linewidth]{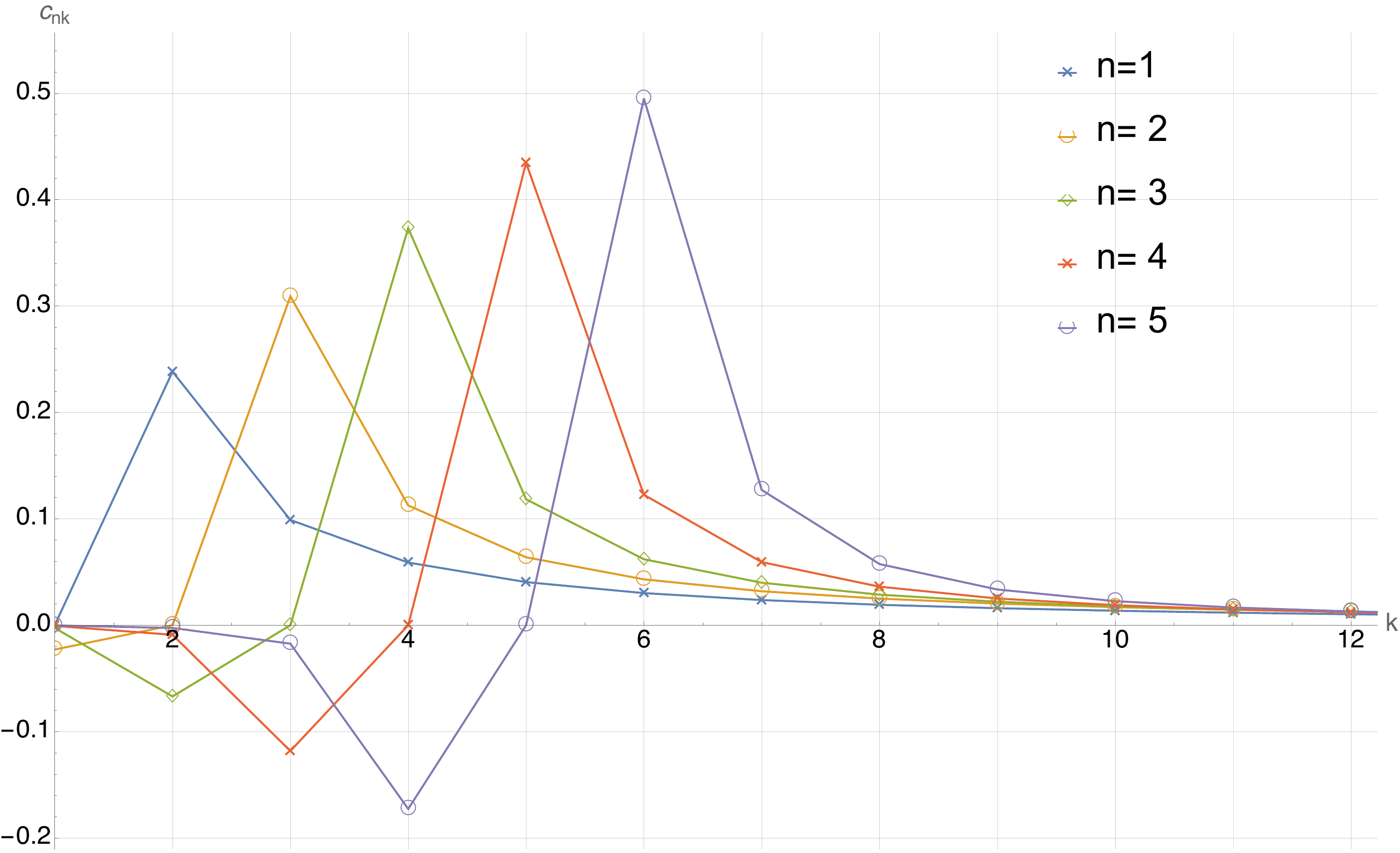}
    \caption{$c_{nk}$ for $n=1,2,3$ and $k \in [1,12]$.  $c_{nk}$ for $n=1,2,3,5,5$ and $k\in[1,12]$. Lines connect discrete values of $c_{nk}$ to enhance visualization of how $c_{nk}$ varies with discrete changes in $n$ and $k$.}
    \label{Fig:c1k}
\end{figure}
\begin{eqnarray}
    \psi &=& \psi^{(0)} + \epsilon \psi^{(1)} + O(\epsilon^2), \label{psi_series}\\
    E &=& E^{(0)} + \epsilon E^{(1)} + O(\epsilon^2), \label{energy_series}\\
    V &=& V^{(0)} + \epsilon V^{(1)} + O(\epsilon^2) \label{potential_series}
\end{eqnarray}
The zero-order potential \( V^{(0)} \) is determined by the expression:
\begin{equation}
\label{V0}
V^{(0)} = -\frac{q}{4\pi  r},
\end{equation}
leading to the following zero-order Schr\"odinger equation for the hydrogen-like atom:
\begin{eqnarray}
\left(-\frac{\nabla^2}{2m} - \frac{q^2}{4\pi  r}\right) \psi^{(0)} &=& E^{(0)} \psi^{(0)}.
\end{eqnarray}
Let it be highlighted that $\psi^{(1)}$ is solved by:
\begin{equation}
    \left(-\frac{\nabla^2}{2m}-\frac{q^2}{4\pi  r}\right) \psi^{(1)} = (E^{(1)}- q V^{(1)})\psi^{(0)}.
\end{equation}
Let us represent the energy state $n$ by $\psi_n(r)$, it has the following 
$\epsilon$ expansion:
\begin{eqnarray}
\label{psi_n_ref}
    \psi_n(r) &=& \psi_n^{(0)}(r) + \epsilon \psi_n^{(1)}(r) + O(\epsilon^2).
\end{eqnarray}
A complete set of basis functions for the spherical zero-order solution is given by:
\begin{equation}
\label{psi0n}
\psi^{(0)}_n(r) = \sqrt{\frac{(n-1)!}{\pi a_0^3 n^4 n!}} \exp\left(-\frac{\rho}{2}\right) L^{(1)}_{n-1}\left(\rho\right)
\end{equation}
where \(\rho = \frac{2r}{n a_0^*}\), and \(a_0^*\) is the modified Bohr radius defined as:
\(
a_0^* = \frac{4\pi}{m q^2}.
\)
In this formulation, \(L^{(1)}_{n-1}(\rho)\) denotes the associated Laguerre polynomial of degree \(n-1\), and \(n\) represents the principal quantum number. The zero-order energy of state $n$ is given by 
\begin{equation}
\label{E0Ref}
    E^{(0)}_n= -\frac{q^2}{8\pi a_0^* n^2}=-\frac{1}{2m {a_0^*}^2 n^2}.
\end{equation}
In order to solve the equations at the first order of perturbation, we note that the first-order potential correction satisfies:
\begin{eqnarray}
    \nabla^2 V^{(1)} = q |\psi^{(0)}|^2.
\end{eqnarray}
The total electron's charge confined within a radius \( r \) is given by:
\begin{equation}
\label{QrRef}
    Q^{(1)}[\psi^{(0)}] = 4\pi q \int_0^r r'^2 |\psi^{(0)}(r')|^2 \, dr'.
\end{equation}
This allows us to calculate the first correction to the electric potential by integrating the charge distribution:
\begin{eqnarray}
\label{V1Ref}
    V^{(1)}[\psi^{(0)}](r) = - \int_r^\infty \frac{Q[\psi^{(0)}](r')}{4\pi {r'}^2} \, dr',
\end{eqnarray}
where the integration is taken from \( r \) to infinity. Notice that $Q^{(1)}[\psi^{(0)}]$ and $V^{(1)}[\psi^{(0)}]$ are functional of $\Psi^{(0)}$. For sake of simplicity, when needed, we will present them by $Q^{(1)}$ and $V^{(1)}$. 

Note that both the zeroth-order wave function, \(\psi^{(0)}\), and the perturbed wave function, \(\psi\), are normalized to unity. Therefore, we have:
\[
\int d^3 x \, \psi^{(0)*} \psi^{(1)} = 0.
\]
This implies that the first-order wave function, \(\psi^{(1)}\), is orthogonal to the zeroth-order wave function under the normalization condition. Multiplying both sides of Equation \eqref{perturb_sch} by \(\psi^{(0)*}\) and integrating over the entire space yields:
\begin{equation}
\label{E1Ref}
E^{(1)} = 4\pi q \int_0^\infty r^2 dr \, V^{(1)}[\psi^{(0)}] \, |\psi^{(0)}|^2,   
\end{equation}
which represents the first-order correction to the total energy. The zero-order total energy is $M+m+ E^{(0)}$, and $E^{(1)}$ always remain perturbative.  

When $\psi^{(0)}= \psi_n^{(0)}$, the standard perturbation theory also leads to:
\begin{eqnarray}
\label{psi_n_1}
    \psi^{(1)} &=& \sum_{k\neq n} c_{nk} \psi_k^{(0)}\\
    c_{nk}  &=&\frac{q  }{E_n^{(0)}-E^{(0)}_k} \int d^3x \psi^{(0)*}_k \psi^{(0)}_n  V^{(1)}[\psi^{(0)}_n]\, 
\end{eqnarray}
Table \ref{tab:values} presents the first-order corrections to the potential, while Figure \ref{Fig:c1k} illustrates the coefficients \( c_{nk} \). We observe that \( c_{nk} \) rapidly decreases for \( k > n \), allowing us to truncate the series and perform the computations numerically. Additionally, we note that \( |c_{nk}| \) may not remain below $1$ for all values of \( n \) and \( k \). This observation may challenge our postulate that treating \( \epsilon \) as a perturbation enhances predictive accuracy. However, we also find that such violations occur only for a few \( k \) values within each \( n \). This suggests that the perturbative approach may still yield approximate results despite these limited instances of non-negligible \( c_{nk} \).
\begin{table}[t!]
\centering
\begin{tabular}{|c|c|c|}
\hline
$n$ & $E_n^{(1)}$ &$\frac{4\pi a_0 n^2 \rho}{q} e^{-\rho} V^{(1)}_n$ \\
\hline
1 & $-\frac{5 q^2}{32 a_0^* \pi }$ &$\rho - 2 e^{\rho} + 2$ \\
2 & $-\frac{77 q^2}{2048 a_0^* \pi }$ &$\frac{\rho^3}{2} + \rho^2 + 3 \rho - 4 e^{\rho} + 4$ \\
3 & $-\frac{17 q^2}{1024 a_0^* \pi }$&$\frac{\rho^5}{12} - \frac{\rho^4}{6} + \rho^3 + 2 \rho^2 + 5 \rho - 6 e^{\rho} + 6$ \\
\hline
\end{tabular}
\caption{The first correction to $E$ of the electron and the electric potential for the first lowest energy state of the electron $n$ where $r=\frac{n a_0 \rho}{2}$.}
\label{tab:values}
\end{table}

\subsection{Modeling the Proton}
\label{sec:ProtonModel}

Protons, though fundamentally composite particles composed of quarks and gluons bound by the strong nuclear force, are often approximated by simpler models when examining atomic electron spectra. Such approximations are justified because electron energies in hydrogen atoms predominantly depend on electromagnetic interactions at scales much larger than the proton's internal quark-gluon dynamics. In our analysis, we adopt a simplified yet physically motivated model, representing the proton as a uniformly charged sphere with a finite radius \(a\). This model assumes a homogeneous and isotropic distribution of energy density, akin to a perfect fluid configuration.

In this simplified scenario, the proton's total positive charge \(+q\) is uniformly distributed on its spherical surface, generating an outward-directed electrostatic repulsion among the charges. To achieve mechanical equilibrium and maintain stability, this electrostatic repulsion must be balanced by an inward-directed internal pressure. The internal pressure arises naturally from the proton’s internal structure governed by quantum chromodynamics (QCD), though here it is modeled phenomenologically.
Consider a small segment of the proton’s charged surface: the electrostatic force per unit area (pressure) acting outward on this segment is given by:
\begin{equation}
\frac{\sigma^2}{2}, \quad \text{where} \quad \sigma = \frac{q}{4\pi a^2}
\end{equation}
represents the surface charge density. Since external pressures are negligible, the proton’s internal pressure \(p\) must precisely counteract this electrostatic repulsion, yielding:
\begin{equation}
    \label{p_def}
p = -\frac{\sigma^2}{2} = -\frac{q^2}{32\pi^2 a^4}.
\end{equation}
The negative sign indicates that the pressure is inward-directed, characteristic of confining forces such as the strong nuclear force. Within this simplified spherical model, we define the proton’s energy density \(\rho\) uniformly as:
\begin{equation}
    \label{rho_def}
    \rho = \frac{3M}{4\pi a^3},
\end{equation}
where \(M\) is the proton’s mass.

Although straightforward, this finite-size proton model introduces subtle corrections to the hydrogen atom’s energy levels, most notably affecting electrons in states with zero orbital angular momentum (S-states). These electrons have a finite probability density at the nucleus, where differences from a point-like proton potential are most significant. Consequently, energy level shifts arise, albeit small, typically on the order of fractions of a percent relative to the unperturbed energy.

While our simplified model adequately captures the primary electromagnetic interactions relevant to the hydrogen electron spectrum, it inherently neglects finer details such as non-uniform charge distributions, the proton’s magnetic moment contributions (which induce hyperfine structure), and dynamic internal quark-gluon effects fully described by QCD. Nonetheless, these effects are substantially smaller than those considered here, justifying their omission in this context.

\section{Energy Momentum tensor}
\label{sec:EMTensor}
The energy momentum tensor of the hydrogen atom can be represented as the sum of the contributions from the electron, the proton, and the electromagnetic field. This is expressed as:
\begin{equation}
    T_{\mu\nu} = T^{\gamma}_{\mu\nu} + T^{e}_{\mu\nu} + T^{p}_{\mu\nu}.
\end{equation}
In the following sections, we will calculate the energy momentum tensors of the proton, the electron, and the electromagnetic field within the hydrogen atom.

The energy-momentum tensor \( T^{\mu\nu} \) characterizes the distribution of energy, momentum, and stress within a field or medium in the framework of general relativity and field theory. For a perfect fluid, this tensor incorporates both the energy density \(\rho\) and the isotropic pressure \(p\), essential for describing fluid dynamics in spacetime. The formal expression is given by:
\[
T^{\mu\nu} = (\rho+ p)  u^{\mu} u^{\nu} + p g^{\mu\nu}.
\]
Here, \(u^{\mu}\) represents the four-velocity of the fluid elements, indicative of their displacement through time and space, and \(g^{\mu\nu}\) denotes the metric tensor. In the rest frame of the proton we have \(u^{\mu}=(1, 0,0,0)\), therefore in the Cartesian coordinates we get:
\begin{eqnarray}
T_{00}^{p} &=&  \rho,\\
T_{ij}^p &=& p \delta_{ij}, 
\end{eqnarray}
where $\rho$ and $p$ are given in \eqref{rho_def} and \eqref{p_def}, and the superscript $p$ stands for proton. Outside the proton, $T_{\mu\nu}^{p}=0$. This leads to:
\begin{eqnarray}
    \int d^3x T_{00}^p &=& M, \\
    \int d^3x T_{ij}^p &=& -\frac{q^2}{24 \pi a} \delta_{ij},
\end{eqnarray}
where $M$ is the mass of the proton and $a$ is its radius. 
\begin{table}[t!]
\centering
\begin{tabular}{|c|c|c|}
\hline
\text{Configuration} & ${\cal E}_n^{\text{NIST}}$ &\text{Error}$(\text{cm})^{-1}$ \\
\hline
1S & 0 & 0.0000000010 \\
2S & 82258.9543992821      & 0.0000000010 \\
3S & 97492.221701 & 0.000007\\
4S & 102823.8530211 & 0.0000003 \\
5S & 105291.63094& 0.00004 \\
6S &  106632.1498416&0.0000007 \\
7S &107440.43933& 0.00004 \\
8S & 107965.0497145& 0.0000003 \\
9S & 108324.72055& 0.00004\\
10S &108581.99080 & 0.00004 \\
11S & 108772.34157&  0.00004 \\
12S & 108917.11886& 0.00004\\
\hline
\end{tabular}
\caption{The reported experimental atomic spectra in $(\text{cm})^{-1}$ for spherical configuration of electron in the hydrogen atom.}
\label{tab:NIST}
\end{table}

The energy momentum tensor of matter action is defined by:
\begin{eqnarray}
    T_{\mu\nu} &=&  2 \frac{\partial {\cal L}_{matter}}{\delta g^{\mu\nu}} - g_{\mu\nu} {\cal L}_{matter}.
\end{eqnarray}
Here, \(\mathcal{L}_{\text{matter}}\) signifies the Lagrangian component exclusive of gravitational interactions. In this section, the calculation is executed under Cartesian coordinates with the metric expressed as:
\(
ds^2 = dx_0^2 - \sum dx_i^2
\). Focusing on a spherical solution, the non-zero components of the energy-momentum tensor are:
\begin{eqnarray}
    T^{e}_{00} & = &  2 |D_0 \phi|^2 - {\cal L}_e, \\
    T^{e}_{ii} & = & 2 |\partial_i \phi|^2 + {\cal L}_e.
\end{eqnarray}
Applying the solution \(\phi\) from equation \eqref{phi_sol} and considering low-energy limits, we obtain:
\begin{eqnarray}
    T^e_{00} &=& (m + E - q V)|\psi|^2 + \frac{|\nabla \psi|^2}{2m},\\
    T^e_{ii} &=&  (E- q V)|\psi|^2 - \frac{|\nabla \psi|^2}{2m} + \frac{|\partial_i \psi|^2}{m},
\end{eqnarray}
where \(|\nabla \psi|^2 = \partial_i \psi^* \partial^i \psi\). Integrating over whole of space and utilizing the equation of motion and the spherical symmetry yield:
\begin{eqnarray}
    \int d^3x\, T^{e}_{00} & = & m, \\
    \int d^3x\, T^{e}_{ii} & = & \frac{4}{3}(E- e \overline{V}),
\end{eqnarray}
where 
\(\overline{V}= \int d^3 x |\psi|^2 V\).  The energy-momentum tensor for the electrodynamics scenario, specifically from the gauge field dynamics described in \eqref{A_mu}, is:
\begin{eqnarray}
    T_{00}^\gamma &=& \frac{1}{2} |\vec{E}|^2 - \rho_P V,\\
    T_{ii} ^\gamma &=&\frac{1}{2} |\vec{E}|^2+ E_i^2 - \rho_p V,
\end{eqnarray}
where \(\rho_p\) denotes the charge density of the proton, \( \vec{E} = -\nabla V \) represents the electric field, and \(E_i\) is the \(i\)-th component of \(\vec{E}\) in the coordinate \(x^i\). Spherical symmetry implies 
\(
    \int d^3x E_i^2 = \frac{1}{3} \int d^3x |\vec{E}^2|
\)
Utilizing the proton model, we deduce:
\begin{eqnarray}
    \int d^3x T_{00}^\gamma &=& \frac{1}{2}\int d^3x |\vec{E}|^2 +q V(a),\\
    \int d^3x  T_{ii} ^\gamma &=&\frac{1}{6} \int d^3x |\vec{E}|^2 - q  V(a).
\end{eqnarray}
Incorporating the identity of
\(\int d^3 x |E|^2 = - \int d^3x V \nabla^2 V\),
and utilizing \eqref{v_eq1}, we then find:
\(\int d^3x |E|^2 = -q V(a)+ q \overline{V}\).
This leads to the final expressions for the integrals of the tensor components:
\begin{eqnarray}
    \int d^3x T_{00}^\gamma &=&\frac{1}{2} q  (\overline{V}+ V(a))\\
    \int d^3x  T_{ii} ^\gamma &=&\frac{1}{6} q (\overline{V}- 7 V(a))
\end{eqnarray}

\section{Spectra of Hydrogen Atom}
\label{sec:Spectra}
\begin{figure}[t]
    \centering
    \includegraphics[width=.90\linewidth]{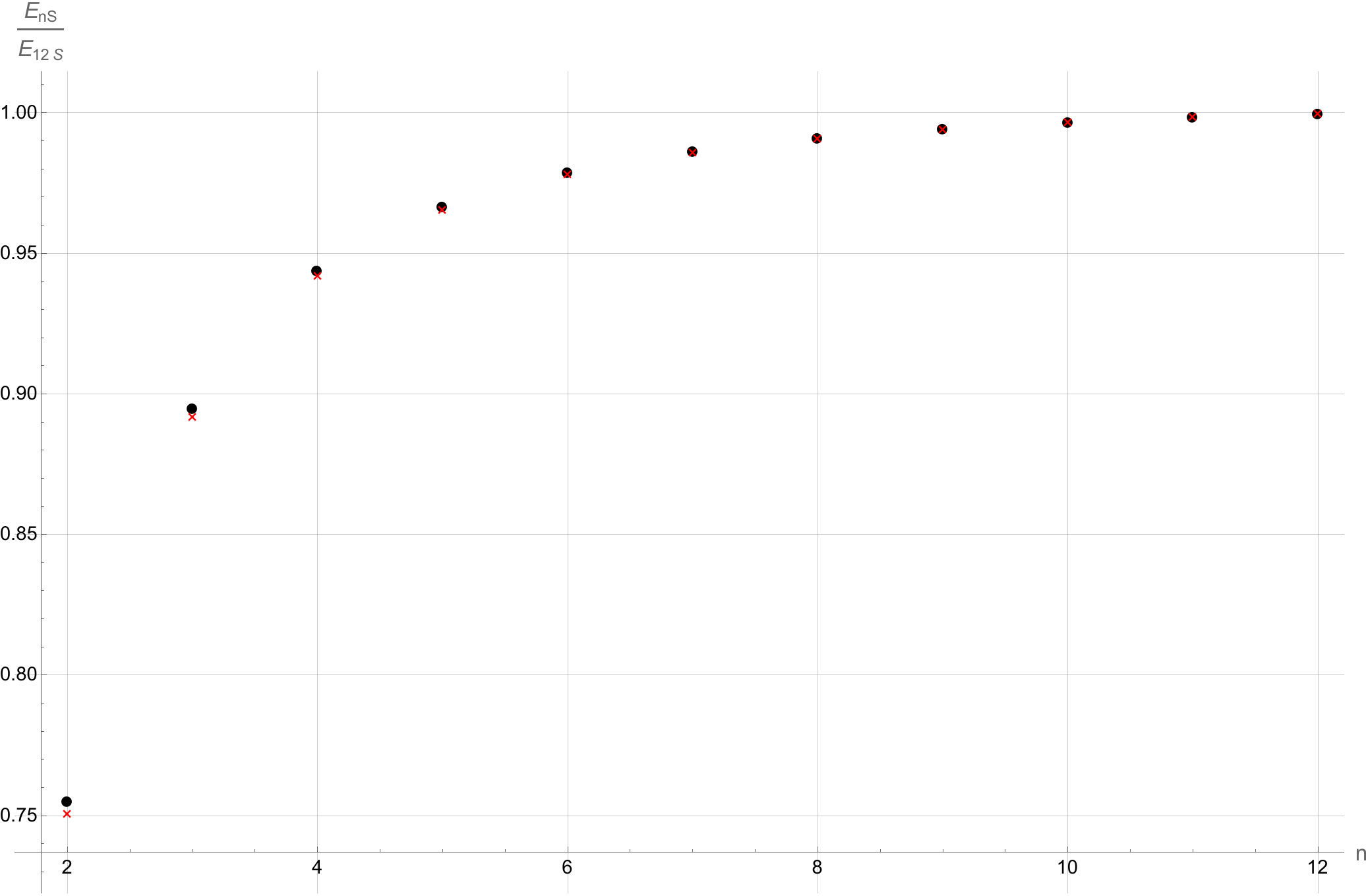}
    \caption{ Dimensionless spectral values for the hydrogen atom's nS states are shown, comparing theoretical predictions (red; see Eq. \ref{Entheory}) with experimental measurements (blue; see Eq. \ref{NistReport})}
    \label{Fig:NistTheory}
\end{figure}

\begin{figure}[t]
    \centering
    \includegraphics[width=.90\linewidth]{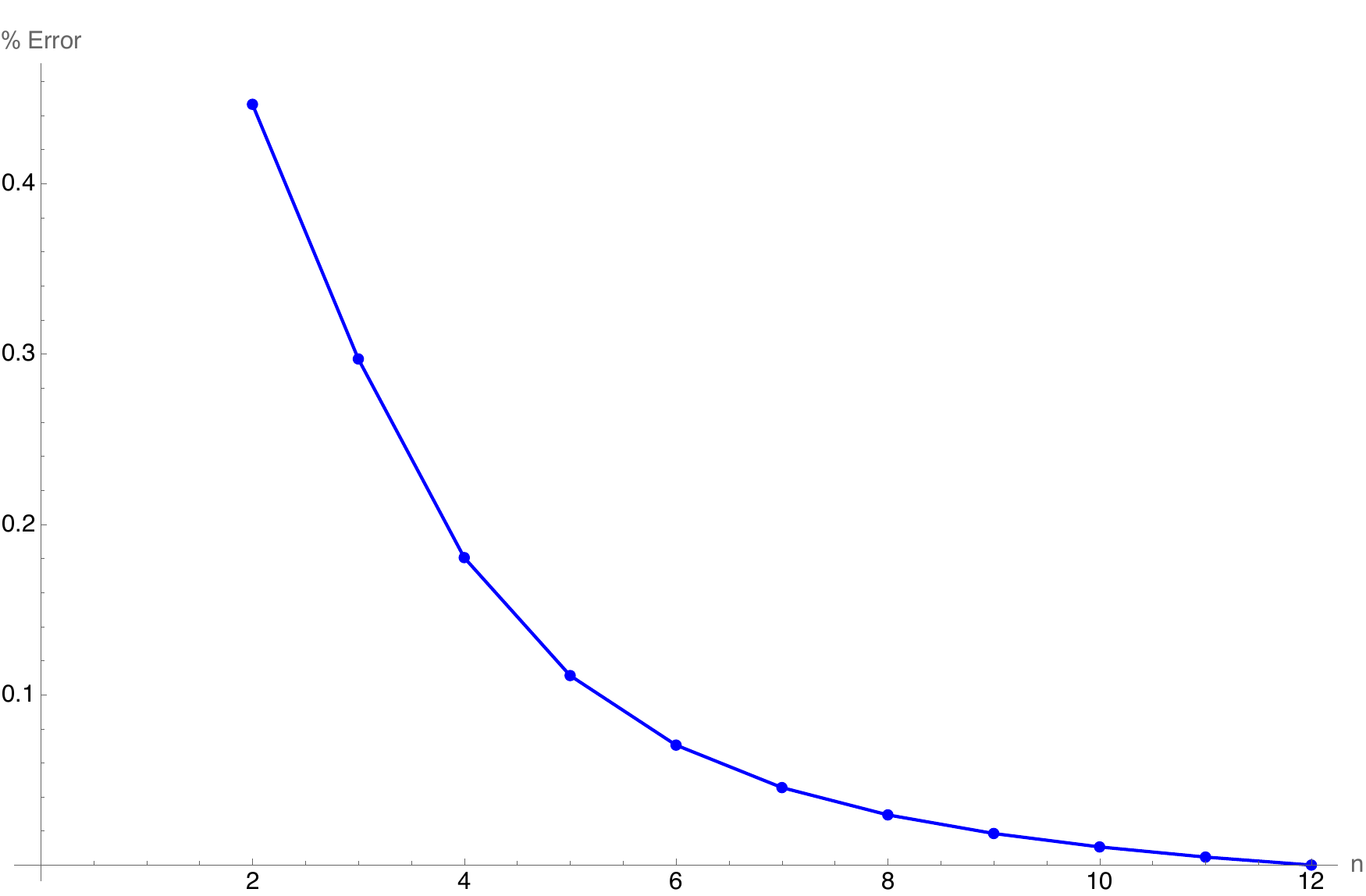}
    \caption{ Percent error between the experimental and theoretical energy differences between state $n$ and state $1$, normalized by the energy difference between states $1$ and $12$.}
    \label{Fig:Error}
\end{figure}

The non-zero spatial integral of the components of the total energy momentum of the hydrogen atom are given by:
\begin{subequations}
\label{Tcomponents}
\begin{align}
    \int \mathrm{d}^3x \, T_{00} &= M + m + \frac{1}{2} q \left( \overline{V} + V(a) \right), \\
    \int \mathrm{d}^3x \, T_{ii} &= \frac{4}{3} E - \frac{q^2}{4\pi a} - \frac{7}{6} q (\overline{V} + V(a)).
\end{align}
\end{subequations}
So the Energy of the hydrogen atom in its rest frame when electron has the wave function of $\psi$, as defined by $\int d^3x T_{00}- M - m$, is given by:
\begin{equation}
\label{EnergyPsi}
    {\cal E} = \frac{1}{2} q(\overline{V} +  V(a))
\end{equation}
Notice that $\overline{V}$ and $V(a)$ depend on the electron's wave function.
Eq. \eqref{EnergyPsi} identifies the total energy of the hydrogen atom in state $\psi$. The energy difference between state $12S$ and $nS$ is given by: 
\begin{equation}
\label{EnTheory}
    {\cal E}_n^{theory} = \frac{q}{2} \left(\overline{V[\psi_n]} - \overline{V[\psi_{12}]}+V[\psi_n](a)-V[\psi_{12}](a)\right)
\end{equation}
The wave function of state $nS$ is given by 
\begin{equation}
    \psi_n= \psi_n^{(0)} + \epsilon \psi_n^{(1)} + O(\epsilon^2)
\end{equation}
where $\psi_n^{(0)}$ is provided in \eqref{psi0n}, and $\psi_n^{(1)}$ is provided in \eqref{psi_n_1}. Table \ref{tab:values} shows the corrections to the first correction to the potential for the first three $nS$ states. So ${\cal E}_n^{theory}$  can be calculated for each $n$. 
The experimental atomic spectra for spherical electron configurations in the hydrogen atom, sourced from the NIST Atomic Spectra Database \cite{NIST_ASD} and detailed in \cite{KRAMIDA2010586}, are presented in Table \ref{tab:NIST}. These values represent the energy differences between the $1S$ state and various $nS$ states. We define the normalized energy ratio as follows:
\begin{eqnarray} 
\label{NistReport}
{\overline{\cal E}}_{n}^{\text{NIST}} = \frac{{\cal E}_n^{\text{NIST}}}{{\cal E}_{12}^{\text{NIST}}} 
\end{eqnarray}
Similarly, the theoretical value of this ratio is calculated using:
\begin{eqnarray} 
\label{Entheory} 
\overline{\cal E}_n^{\text{theory}} = \frac{{\cal E}_n^{\text{theory}}}{{\cal E}_{12}^{\text{theory}}} 
\end{eqnarray}
We have chosen to normalize the ratios with respect to the $12S$ state because relativistic corrections for the $12S$ state are minimal, ensuring a more accurate comparison between experimental and theoretical values. Fig. \ref{Fig:NistTheory} depicts the theoretical versus experimental values. Fig \ref{Fig:Error} depicts 
\(
\frac{ 
    \overline{\cal E}_{n}^{\text{NIST}} 
    - \overline{\cal E}_n^{\text{theory}} 
}
{
    \overline{\cal E}_{n}^{\text{NIST}}
}
\). The average error is $(0.13 \pm 0.04) \%$.  

\section{Linear Momentum}
\label{sec:LinearMomentum}
In this section, we analyze the linear momentum of the hydrogen atom under a Lorentz boost in the $z$-direction with velocity $v$. The corresponding Lorentz transformation is given by
\begin{equation}
\Lambda =
\begin{pmatrix}
 \gamma & 0 & 0 & \beta\gamma \\
 0      & 1 & 0 & 0 \\
 0      & 0 & 1 & 0 \\
 \beta\gamma & 0 & 0 & \gamma
\end{pmatrix},
\end{equation}
where $\gamma = 1/\sqrt{1-\beta^2}$ and $\beta = v/c$. Under this transformation, the energy-momentum tensor in the boosted frame, $T'_{\mu'\nu'}$, is related to the tensor in the rest frame, $T_{\mu\nu}$, by
\begin{equation}
T'_{\mu'\nu'} = \Lambda^{\mu}_{\ \mu'}\, \Lambda^{\nu}_{\ \nu'}\, T_{\mu\nu}\,.
\end{equation}

Of particular interest is the $03$-component of the energy-momentum tensor. After performing the spatial integration, one obtains
\begin{equation}\label{eq:QFT_momentum}
\frac{1}{\beta\gamma}\int d^3x'\, T'_{03} =  M + m - \frac{q^2}{4\pi a} + \frac{4}{3} (E - {\cal E}),
\end{equation}
where the relation given in \eqref{EnergyPsi} has been used. In particular, this result is achieved without resorting to an $\epsilon$-expansion. Notice that $E$ is the time oscillation in the wave function while $\cal E$ is the bound-state energy of the electron. 

Now let us consider the initial setup where we have a hydrogen atom in the \( nS \) state moving with velocity \( v \) along the positive \( z_3 \) direction, while another hydrogen atom in its ground state is moving with the same speed \( v \) along the negative \( z_3 \) direction. According to \eqref{eq:QFT_momentum}, the linear momentum of the system is given by
\begin{equation}
    p = \frac{4v\gamma}{3}\left(1 - \frac{E_n - E_1}{{\cal E}_n - {\cal E}_1}\right)({\cal E}_n - {\cal E}_1).
\end{equation}
Since we are working within a renormalizable theory, the theoretical energy differences are mapped onto their experimental counterparts. Therefore, we can write
\begin{eqnarray}
\label{Qp}
        p &=& \frac{4v\gamma}{3}\alpha_n {\cal E}_n^{\text{NIST}},\\
        \alpha_n &=& 1 - \frac{E_n - E_1}{{\cal E}_n^{\text{theory}}}
\end{eqnarray}
where \({\cal E}_n^{\text{NIST}}\) is the experimentally observed energy difference between the \(1S\) and \(nS\) states, and \({\cal E}_n^{\text{theory}}\) is given in \eqref{EnTheory}. With these expressions, we can proceed to calculate \({\cal E}_n^{\text{theory}}\) and \(E_n - E_1\), and identify $\alpha_n$. Table \ref{tab:alpha_n} shows the expansion series $\epsilon$ for $\alpha_n$. We observe that at the leading order, $\alpha_n=1$, and \eqref{eq:relativistic_momentum} and \eqref{Qp} differs with a factor of $\frac{4}{3}$. This proves that  the origin of the 4/3 factor is not due to extra, hidden momentum associated to poincare stress or other form of energy but rather to the nonlocality (deviation from point-like particle) inherent in the electron's distribution.

Table \ref{tab:alpha_n} displays the first perturbative correction to 
$\alpha_n$, which in turn modifies the 4/3 factor. To our knowledge, this effect has not been reported in previous literature. In particular, the leading correction at $\epsilon =1$ is substantial; it effectively reduces the factor from 4/3 to 2/3. This sizable correction suggests that higher-order terms could play an important role; however, their detailed analysis is beyond the scope of this work. The large first correction suggests that the perturbation series may be alternating or slowly convergent. It is conceivable that including the next terms might bring the factor closer to unity (the expected $E/c^2$), but a non-perturbative summation might be required to confirm this. 
\begin{table}[t]
\label{tab:alpha_n}
\begin{ruledtabular}
\begin{tabular}{cc}
$n$ & $\alpha_n$ \\ \hline
2 & $1-0.481969\,\epsilon + O\left(\epsilon^2\right)$ \\
3 & $1-0.49793\,\epsilon + O\left(\epsilon^2\right)$ \\
4 & $1-0.504835\,\epsilon + O\left(\epsilon^2\right)$ \\
5 & $1-0.508321\,\epsilon + O\left(\epsilon^2\right)$ \\
6 & $1-0.510261\,\epsilon + O\left(\epsilon^2\right)$ \\
7 & $1-0.511425\,\epsilon + O\left(\epsilon^2\right)$ \\
8 & $1-0.512169\,\epsilon + O\left(\epsilon^2\right)$ \\
9 & $1-0.512675\,\epsilon + O\left(\epsilon^2\right)$ \\
10 & $1-0.513037\,\epsilon + O\left(\epsilon^2\right)$ \\
11 & $1-0.513308\,\epsilon + O\left(\epsilon^2\right)$ \\
12 & $1-0.513517\,\epsilon + O\left(\epsilon^2\right)$ \\
\end{tabular}
\end{ruledtabular}
\caption{Series Expansion Coefficients $\alpha_n$. }
\end{table}

\section{Conclusion}
\label{sec:Conclusion}

In this work, we have revisited the long-standing electromagnetic mass problem from a modern quantum field theoretic perspective. By analyzing a system of two widely separated hydrogen atoms---one in an excited \(nS\) state and the other in the ground \(1S\) state---we isolated the electromagnetic contribution to the total linear momentum and the associated mass of the electron. Our approach circumvents the need to invoke ad hoc constructs such as Poincaré stresses or the rigid-body assumptions of earlier models.

Using a Lorentz-boosted energy–momentum tensor, we computed the linear momentum of a moving hydrogen atom. Our analysis revealed a significant discrepancy between the momentum obtained via the full quantum field theoretic treatment and that predicted by the point-like particle approximation. Notably, the leading perturbative correction introduces a factor of \(4/3\), which, along with the evaluated subleading corrections, indicates that the effective electromagnetic mass is not simply given by \(E/c^2\). Instead, these corrections underscore the role of nonlocality in the electromagnetic field configuration. The substantial size of the leading correction at \(\epsilon=1\)---effectively modifying the \(4/3\) factor toward \(2/3\)---suggests that higher-order terms may further influence the momentum balance. Although a comprehensive analysis of these higher corrections lies beyond the scope of the present study, our results strongly imply that nonlocal (non-pointlike) effects are intrinsic to the electromagnetic field's contribution to the total momentum.

A further intriguing aspect of our study emerges when comparing the predictive power of the Schrödinger equation with that of our quantum field theory (QFT) approach. Using the Schrödinger equation, we achieve an average error of $(1.4 \pm 0.2) \times 10^{-5}\%$ between theoretical predictions and experimental measurements. To enhance the agreement between theory and experiment, we incorporated the self-interaction of the electron's wave function—neglected by the Schrödinger equation—as perturbations within a robust QFT framework. Contrary to our expectations, this approach did not improve the results; instead, it increased the error to $(0.13 \pm 0.04)\%$, a behavior also reported in \cite{Biguaa:2020omt}.

This counterintuitive outcome, where adding supposedly necessary corrections worsens the prediction, leads to the question of why the linear Schrödinger equation achieves such high accuracy despite neglecting explicit self-interaction. The answer likely lies in the fact that the physical electron mass used in the Schrödinger equation is already an empirical, renormalized mass. This parameter implicitly incorporates the electromagnetic self-energy effects that our perturbative QFT treatment attempted to add explicitly. In standard quantum mechanics, the electron mass effectively absorbs these self-interaction contributions through renormalization. Consequently, the Schrödinger equation benefits from this built-in correction, avoiding the double counting or improper application of self-energy effects that seems to occur in our perturbative QFT approach, thereby achieving remarkable accuracy despite its apparent simplicity.

During our analysis, we made two primary postulates:
\begin{enumerate}
\item \textbf{Renormalizability Ensures Spectral Accuracy}: The renormalizability of the theory guarantees the correct identification of all spectral features once the energy differences between any two states are determined.
\item \textbf{Perturbative Enhancement of Predictive Accuracy}: Treating $\epsilon$ in \eqref{perturb_pot} as a perturbation will enhance the predictive accuracy of our model.
\end{enumerate}
We believe the first postulate holds firm. However, the failure of our perturbative QFT approach to improve accuracy, as demonstrated by the increased error and further suggested by the fact that coefficients $c_{nk}$ (Figure \ref{Fig:c1k}) do not remain negligible, indicates that the second postulate requires reconsideration. The perturbative expansion in $\epsilon$ did not enhance predictive accuracy as expected. This suggests that simply adding these terms perturbatively is insufficient or incorrectly applied, possibly due to the complex interplay already captured by the renormalized mass in the simpler theory. A more robust QFT solution might necessitate accounting for all $\epsilon$ expansions to find a non-perturbative solution or employing different techniques altogether. Addressing the limitations of the current perturbative approach may involve exploring alternative perturbative schemes, incorporating relativistic corrections more fully, or developing non-perturbative methods to better capture the underlying physics. Further investigation is essential to understand why this perturbative QFT correction scheme fails where the Schrödinger equation, implicitly corrected via renormalization, succeeds. Addressing this issue may involve exploring alternative perturbative techniques, incorporating relativistic corrections, or developing non-perturbative methods to better capture the underlying physics. Further investigation is essential to understand the limitations of the current approach and to identify strategies to improve the theoretical predictions.

In summary, our study provides compelling evidence that the origin of the \(4/3\) factor—and the corresponding mismatch in linear momentum—is a direct consequence of the non-local nature of the electromagnetic field. Future work aimed at resolving the discrepancies between the QFT and Schrödinger approaches may benefit from exploring non-perturbative techniques such as directly solving the coupled Maxwell-Dirac (or Maxwell-Klein-Gordon) equations for bound states, employing lattice gauge theory, or utilizing variational methods and the Bethe-Salpeter formalism\cite{PhysRev.84.1232}. Such avenues promise not only to further elucidate the electromagnetic mass problem but also to enhance our broader understanding of bound-state dynamics in quantum field theory.


\begin{thebibliography}{20}
\bibitem[Not()]{Note}
	There are three distinct definitions of mass commonly discussed in physics. The
	first arises from Newton’s second law of motion and is associated with
	inertia—this is inertial mass, which quantifies an object’s resistance to
	acceleration when a force is applied. The second definition comes from
	gravitational interactions and is known as gravitational mass—it determines
	the strength of the gravitational force an object experiences in a
	gravitational field. The third definition connects mass to energy through
	Einstein’s theory of relativity, particularly via the famous relation $E =
	mc^2$, and is referred to as relativistic or energy-equivalent mass. Although
	these three definitions stem from different physical principles and
	theoretical frameworks, no experimental evidence to date has demonstrated any
	discrepancy among them. All available observations and measurements suggest
	that inertial mass, gravitational mass, and the mass-energy equivalence yield
	consistent results, reinforcing the idea that mass is a coherent and unified
	concept across different domains of physics.
\end{thebibliography}
\end{document}